\documentclass[aps,reprint,amsmath,amssymb,groupaddress,superscriptaddress,prx]{revtex4-2}

\usepackage[utf8]{inputenc}
\usepackage{amsmath}
\usepackage{amsfonts}
\usepackage{amssymb}
\usepackage{physics}
\allowdisplaybreaks[4]
\usepackage{bm}
\usepackage{color}
\usepackage{graphicx}
\usepackage{soul}
\usepackage{CJK}
\usepackage{xcolor}
\usepackage{mathtools}
\usepackage{lipsum}
\usepackage{comment}

\usepackage{xcolor}
\usepackage[colorlinks,citecolor=blue,linkcolor=blue,anchorcolor=blue,urlcolor=blue]{hyperref}
\usepackage{chngcntr}
\begin{document}

\title{Finite-time thermodynamics: A journey beginning with optimizing heat engines}

\author{Yu-Han Ma}
 \email{yhma@bnu.edu.cn}
\affiliation{School of Physics and Astronomy, Beijing Normal University, Beijing, 100875, China}
\affiliation{Graduate School of China Academy of Engineering Physics, Beijing, 100193, China}

\author{Xiu-Hua Zhao}
\affiliation{School of Physics and Astronomy, Beijing Normal University, Beijing, 100875, China}

\date{\today}

\maketitle

\section{Background}

The utilization of fire and the heat it provides guided early humans through millennia of cultural evolution. 
Around three centuries ago, the widespread use of heat engines, a key energy technology innovation during the first industrial revolution, rapidly propelled human civilization into the modern era~\cite{smilEnergyCivilizationHistory2017}. Undoubtedly, the development of thermodynamics plays a crucial role in this process. The first and second laws of thermodynamics together lead to an interesting result: the efficiency of any heat engine converting heat into work is bounded from above by the Carnot efficiency $\eta_{\rm{C}}\equiv 1-T_c/T_h$, where $T_h$($T_c$) is the temperature of the hot (cold) reservoir from which the engine absorbs (releases) heat~\cite{CarnotFrench,carnotEnglish}. $\eta_{\rm{C}}$ serves as the most fundamental constraint in thermodynamics~\cite{callenThermodynamicsIntroductionThermostatistics1991}, determined solely by the temperatures of the reservoirs, independent of the specific properties of the working substance. Classical thermodynamics implies that the reversible thermodynamic cycle required to achieve Carnot efficiency can only be realized in the quasi-static limit. The resulting zero output power makes the Carnot cycle entirely impractical for real-world applications. 

Since the industrial revolution, although all practical heat engines satisfy the constraint of Carnot's theorem, there exists a significant gap between their operational efficiencies and $\eta_{\rm{C}}$~\cite{bejanAdvancedEngineeringThermodynamics2016}. This difference arises from the deviation of actual cycles from the quasi-static Carnot cycle, which cannot be adequately described by equilibrium thermodynamics. Therefore, seeking tighter thermodynamic constraints for practical heat engines that reflect the impact of cycle time is a problem of significant practical value, and resolving this requires new theoretical frameworks that go beyond equilibrium thermodynamics. To address this challenge, finite-time thermodynamics emerges with the core objective of bridging the gap between traditional ideal thermodynamic theories and real-world applications, thereby advancing non-equilibrium thermodynamics and providing quantitative results for near-equilibrium processes~\cite{andresenThermodynamicsFiniteTime1977,bejanAdvancedEngineeringThermodynamics2016}.

\section{Past to current development}
\subsection{Thermodynamic constraint relations}

In the 1950s, French physicist Yvon analyzed the steam cycle in nuclear power plants~\cite{yvonSaclayReactorAcquired1955}. Yvon specified the non-quasi-static feature of the cycle as the heat flow caused by the temperature difference between the high-temperature reservoir and the working substance. Using Newton's law of heat transfer to quantify the heat flow and further identifying cycle power as the optimization objective, he derived the cycle efficiency at maximum power output, now known as efficiency at maximum power (EMP). This new thermodynamic constraint became an important parameter in the later development of finite-time thermodynamics to describe engine performance. Although Yvon's paper was written in French and initially received little attention, his ideas on engine optimization align with many subsequent works in this field. Around the same period, Chambadal~\cite{chambadalCentralesNucleaires1957} and Novikov~\cite{novikovEfficiencyAtomicPower1957} also investigated similar issues separately, obtaining results consistent with Yvon’s. In 1975, Canadian physicists Curzon and Ahlborn proposed a general endo-reversible Carnot cycle model~\cite{curzonEfficiencyCarnotEngine1975}, simultaneously considering the heat flow of the engine in both the high-temperature and low-temperature quasi-isothermal processes. They derived the EMP of the cycle $\eta_{\rm{CA}}=1-\sqrt{T_c/T_h}$, which later became known as Curzon-Ahlborn (CA) efficiency. The CA efficiency received considerable attention from experts in non-equilibrium thermodynamics and engineering thermodynamics~\cite{bejanAdvancedEngineeringThermodynamics2016,berryFiniteTimeThermodynamics2022}, spurring significant development of the emerging field of finite-time thermodynamics over the following decades, and prompting researchers to systematically study the operation of various practical engines within finite-time thermodynamic cycles~\cite{devosEfficiencyHeatEngines1985,sahinEfficiencyJouleBraytonEngine1995,feldmannHeatEnginesFinite1996,chenHeatTransferEffects1998,kaushikFiniteTimeThermodynamic2000}.

Beyond the widely used endo-reversible heat engine models in engineering, researchers have explored the EMP of finite-time heat engines from more fundamental perspectives based on various frameworks of non-equilibrium thermodynamics~\cite{kosloffQuantumMechanicalOpen1984,vandenbroeckThermodynamicEfficiencyMaximum2005,schmiedlEfficiencyMaximumPower2007,tuEfficiencyMaximumPower2008,allahverdyanWorkExtremumPrinciple2008,espositoEfficiencyMaximumPower2010,izumidaEfficiencyMaximumPower2012,shengConstitutiveRelationNonlinear2015,chen2019achieve}. For instance, Van den Broeck applied linear irreversible thermodynamics to analyze the optimization of steady-state heat engines~\cite{vandenbroeckThermodynamicEfficiencyMaximum2005}. Schmiedl and Seifert studied stochastic heat engines with Brownian particles as the working substance, deriving a general expression for the EMP~\cite{schmiedlEfficiencyMaximumPower2007}. Tu explored the finite-time operation of the Feynman ratchet, obtaining its EMP and analyzing the universality of such EMP with respect to Carnot efficiency~\cite{tuEfficiencyMaximumPower2008}. By introducing the $1/\tau$-scaled irreversible entropy generation in finite-time quasi-isothermal processes of duration $\tau$, Esposito et al.~\cite{espositoEfficiencyMaximumPower2010} proved that the EMP of low-dissipation Carnot-like heat engines, $\eta_{\rm{LD}}$, satisfies $\eta_{\rm{C}}/2\leq \eta_{\rm{LD}}\leq \eta_{\rm{C}}/(2-\eta_{\rm{C}})$, where $\eta_{\rm{C}}/2$ and $\eta_{\rm{C}}/(2-\eta_{\rm{C}})$ serve as universal lower and upper bounds of EMP, respectively. At the microscale, Kosloff was the first to investigate the EMP of quantum harmonic oscillator heat engines~\cite{kosloffQuantumMechanicalOpen1984}. Chen et al. ~\cite{chen2019achieve} demonstrated that the extra work of finite-time adiabatic processes (of duration $\tau$) exhibits $1/\tau^2$ scaling, due to which they found that the EMP of a quantum Otto engine can surpass the upper bound, $\eta_{\rm{C}}/(2-\eta_{\rm{C}})$, of finite-time Carnot engines.

The systematic investigation of EMP has also led to another fundamental issue: the potential constraint relation (also called trade-off relation) between power and efficiency, arising from the irreversibility of the cycle~\cite{shiraishi2016universal}. Chen and Yan derived the maximum power for endo-reversible heat engines at a given efficiency~\cite{chenEffectHeatTransfer1989}. Holubec and Ryabov obtained the approximate power-efficiency constraint relation in the low-power region as well as the near-maximum-power region for low-dissipation heat engines~\cite{holubecMaximumEfficiencyLowdissipation2016}. In a recent work~\cite{maUniversalConstraintEfficiency2018}, one of the authors of this paper and collaborators analytically derived a general constraint relation for power and efficiency across the entire parameter space of low-dissipation heat engines:
\begin{equation}
    \frac{1-\sqrt{1-\tilde{P}}}{2}\leq\tilde{\eta}\leq\frac{1+\sqrt{1-\tilde{P}}}{2-\eta_{{\rm C}}\left(1-\sqrt{1-\tilde{P}}\right)},
\end{equation}
where $\tilde{P}$ and $\tilde{\eta}$ are normalized by the maximum power of the heat engine and $\eta_{\rm{C}}$, respectively. Some researchers have also derived constraint relations for power and efficiency using thermodynamic geometry~\cite{abiusoOptimalCyclesLowDissipation2020} and the general dynamical equations of quantum open systems~\cite{brandnerThermodynamicGeometryMicroscopic2020}. Moreover, recent studies have indicated that when considering the fluctuations in the output power of heat engines, there also exists a constraint relation among power, efficiency and power fluctuations~\cite{pietzonkaUniversalTradeOffPower2018,holubecCyclingTamesPower2018,millerThermodynamicUncertaintyRelation2021}. In summary, Carnot efficiency, efficiency at maximum power, and the constraint relation between power and efficiency collectively and progressively characterize the performance of finite-time thermodynamic cycles.

\subsection{Thermodynamic process engineering and optimization}

The boundaries of the aforementioned thermodynamic constraint relation indicate the optimal performance of a heat engine at specific fixed parameters, such as the power-efficiency constraint relation, which represents either the maximum efficiency for a given power or the maximum power for a given efficiency. Achieving these boundaries requires that the operation time or control protocol of the thermodynamic processes satisfy certain conditions~\cite{maOptimalOperatingProtocol2018,chenMicroscopicTheoryCurzonAhlborn2022}. The concept of thermodynamic process control dates back to the study of thermodynamic geometry~\cite{salamonThermodynamicLengthDissipated1983,crooksMeasuringThermodynamicLength2007,sivakThermodynamicMetricsOptimal2012,zulkowskiGeometryThermodynamicControl2012}. In the space defined by thermodynamic state variables, thermodynamic length~\cite{salamonThermodynamicLengthDissipated1983,crooksMeasuringThermodynamicLength2007,sivakThermodynamicMetricsOptimal2012} serves as a metric to quantify the minimum irreversible dissipation that occurs during non-equilibrium processes transitioning between two thermodynamic states~\cite{crooksMeasuringThermodynamicLength2007,sivakThermodynamicMetricsOptimal2012}. Attaining this lower bound requires that the driving of the process adhere to specific criteria, embodying the concept of thermodynamic process control~\cite{li2023nonequilibrium,guery-odelinDrivingRapidlyRemaining2023}. Recently, researchers have extended the near-equilibrium thermodynamic geometry to regions far from equilibrium~\cite{vanvuGeometricalBoundsIrreversibility2021,wangThermodynamicGeometryNonequilibrium2024,zhongLinearResponseEquivalence2024}, enabling broader thermodynamic control.

Physically, the irreversibility of non-equilibrium thermodynamic processes can be quantified by irreversible entropy generation, which, as a process function, depends on the specific driving protocol of the process. Therefore, for a given duration, different driving protocols can lead to varying irreversible entropy generation, resulting in different energy dissipation~\cite{maOptimalOperatingProtocol2018,maExperimentalTestTau2020,nakazatoGeometricalAspectsEntropy2021}. Consequently, to minimize dissipation, researchers have developed optimal control strategies for various thermodynamic processes~\cite{schmiedlOptimalFiniteTimeProcesses2007,gomez-marinOptimalProtocolsMinimal2008,aurellOptimalProtocolsOptimal2011,zulkowskiOptimalControlOverdamped2015,maOptimalOperatingProtocol2018,engelOptimalControlNonequilibrium2023}. Alternatively, the optimization of thermodynamic processes can also consider process time as the optimization objective, aiming for the shortest possible duration under certain constraints. For example, 
when the system interacts with a constant-temperature reservoir, Li et al. proposed a strategy named “shortcuts to isothermality” to realize a finite-time isothermal transition by introducing an auxiliary potential~\cite{liShortcutsIsothermalityNonequilibrium2017,liGeodesicPathMinimal2022}. This approach has been experimentally realized~\cite{albayRealizationFiniterateIsothermal2020} and can be applied to design controllable thermodynamic cycles~\cite{chenOptimizingBrownianHeat2022,zhaoLowdissipationEnginesMicroscopic2022}. Furthermore, several studies have also focused on isothermal shortcuts and the control of thermodynamic cycles for quantum systems~\cite{dannShortcutEquilibrationOpen2019,pancottiSpeedUpsIsothermalityEnhanced2020}.

\subsection{Unconventional heat engines}

After understanding the basic performance constraints and optimal control of various thermodynamic cycles, researchers have started to explore the finite-time performance of unconventional thermodynamic cycles. These studies are primarily motivated by the following two questions: i) In unconventional scenarios, such as limited total energy supply or presence of additional resources, how should heat engines be optimized? ii) How does the performance of a heat engine reflect the non-equilibrium thermodynamic properties of its working substance and the coupled reservoirs? For example, conventional heat engines operate with infinite-sized thermal reservoirs at constant temperature. However, when considering reservoirs of finite size, characterized by finite heat capacity, the operations of the heat engine will cause temperature changes in the reservoir ~\cite{ma2023simple}. In this context, the concept of maximum extractable work has been introduced to describe the performance of the heat engine~\cite{ondrechenMaximumWorkFinite1981}. Correspondingly, the efficiency at maximum work (EMW)~\cite{johalNearequilibriumUniversalityBounds2016} and the efficiency at maximum average power (EMAP) associated with finite-time cycles~\cite{izumidaWorkOutputEfficiency2014} establish the fundamental constraints for heat engines with finite-sized reservoirs. One of the authors of this paper specifically detailed the effects of heat capacity on EMW and EMAP~\cite{maEffectFiniteSizeHeat2020}, and further presented a general constraint relation for the power and efficiency of heat engines~\cite{yuanOptimizingThermodynamicCycles2022} with collaborators. It is worth mentioning that the finiteness of the working substance also affects the efficiency of heat engines~\cite{quan2014maximum,fei2024temperature}, which can be explained by the temperature fluctuations in mesoscopic systems~\cite{fei2024temperature}.

Previously, for the sake of theoretical simplicity, non-interacting systems (such as ideal gases, two-level atoms, and quantum harmonic oscillators) were primarily chosen as working substances in heat engines. However, in the past decade, researchers have started exploring whether many-body interacting systems have thermodynamic advantages for constructing heat engines. It was found that interactions can provide collective advantages, allowing irreversible dissipation to increase more slowly than the available power output as the size of the working substance increases, thus improving the efficiency of the heat engine~\cite{niedenzuCooperativeManybodyEnhancement2018,vroylandtCollectiveEffectsEnhancing2018,latuneCollectiveHeatCapacity2020,filhoPowerfulOrderedCollective2023,rolandiCollectiveAdvantagesFiniteTime2023}. Furthermore, many-body interactions may also induce phase transitions in the working substance, with several studies analyzing the performance enhancement of heat engines through phase transitions~\cite{campisiPowerCriticalHeat2016,holubecWorkPowerFluctuations2017,ma2017quantum}. In Ref.~\cite{liangMinimalModelCarnot2023}, one of the authors of this paper and collaborators proposed a minimal heat engine model with degenerate internal energy levels, which breaks the universal power-efficiency constraint of conventional heat engines, enabling Carnot efficiency at maximum power. Moreover, for micro-scale heat engines, which typically operate in the presence of highly fluctuating energy fluxes, researchers have proposed using fluctuating efficiency to better characterize their performance~\cite{verley2014unlikely,denzler2020efficiency,fei2022efficiency}. When quantum effects are considered, Scully et al. revealed that the quantum coherence of the working substance can enhance engine efficiency~\cite{scullyExtractingWorkSingle2003}, prompting further investigations into leveraging quantum coherence as a thermodynamic resource to optimize heat engine performance~\cite{quan2006maxwell,camatiCoherenceEffectsPerformance2019,ma2021works,fei2022efficiency,aimetEngineeringHeatEngine2023}. 

Recent investigations on active matter have intersected with finite-time thermodynamics, with various researchers proposing the construction of thermodynamic cycles utilizing active matter as working substances or reservoirs~\cite{pietzonkaAutonomousEnginesDriven2019,kumariStochasticHeatEngine2020,leeBrownianHeatEngine2020,fodorActiveEnginesThermodynamics2021,wang2024thermodynamic}. Pietzonka et al.~\cite{pietzonkaAutonomousEnginesDriven2019} proposed a consistent stochastic thermodynamic framework for engines outputting work while being powered by active matter. 
A most recent work by Wang et al.~\cite{wang2024thermodynamic} explored the optimal control of active matter by extending the traditional thermodynamic geometry framework used for passive systems. Furthermore, discussions on heat engines that violate time-reversal symmetry~\cite{benentiThermodynamicBoundsEfficiency2011,brandnerStrongBoundsOnsager2013}, and modifying the interactions between heat engines and thermal reservoirs to regulate the performance of heat engines~\cite{allahverdyanCarnotCycleFinite2013}, also represent a series of beneficial attempts to optimize heat engines. These studies expand the scope of thermodynamic cycle research and provide additional avenues for efficient energy extraction.

\subsection{Experiments}
In contrast to the abundant theoretical advancements, experimental investigations in finite-time thermodynamics are still relatively underdeveloped. In testing the fundamental relations of finite-time thermodynamics, one of the authors of this paper and collaborators utilized an ideal gas platform to measure the irreversible dissipation during the compression process of dry air in contact with a constant temperature reservoir, rigorously verifying the $1/\tau$ scaling of irreversibility in the slow-driving regime~\cite{maExperimentalTestTau2020}. Based on this platform, Zhai et al.~\cite{zhaiExperimentalTestPowerefficiency2023} measure the constraint relation between power and efficiency and EMP in a complete thermodynamic cycle, validating previous theoretical predictions~\cite{vandenbroeckThermodynamicEfficiencyMaximum2005,maUniversalConstraintEfficiency2018,maOptimalOperatingProtocol2018}. Over the past decade, different platforms have successfully implemented various finite-time heat engines and studied their performance. For instance, Blickle and Bechinger realized a microscopic Stirling engine using colloidal particles in an optical trap~\cite{blickleRealizationMicrometresizedStochastic2012}. Ro\ss nagel et al. constructed a single-atom heat engine and measured its efficiency~\cite{rossnagelSingleatomHeatEngine2016}. Mart\'inez et al. designed a Brownian particle Carnot engine and demonstrated its performance advantages~\cite{martinezBrownianCarnotEngine2016}; Krishnamurthy et al. achieved an active Stirling engine operating in a bacterial heat bath~\cite{krishnamurthyMicrometresizedHeatEngine2016}. Besides, some groups have focused on exploring fluctuation relations in micro-scale systems~\cite{liphardtEquilibriumInformationNonequilibrium2002a,wangExperimentalDemonstrationViolations2002,trepagnierExperimentalTestHatano2004,collinVerificationCrooksFluctuation2005,anExperimentalTestQuantum2015,cilibertoExperimentsStochasticThermodynamics2017}, which are also closely related to finite-time thermodynamics.

\section{Future outlook}

Despite substantial research on finite-time thermodynamic processes and heat engine cycles in the near-equilibrium regime, the regime far from equilibrium still requires further exploration. Currently, most research on heat engine cycles concentrates on the slow-driving regime. Although some studies have begun to investigate fast-driving heat engine cycles~\cite{ma2021consistency,cavina2021maximum}, the inherent need for time-scale separation in theoretical tools suggests that more accurate theoretical approaches must be developed. Characterizing the irreversibility of thermodynamic processes in the fast-driving regime~\cite{wang2021nonadiabatic}, as well as optimizing and regulating the performance of heat engines, remain open questions worthy of attention.

Moreover, there has been growing interest in the dynamical processes of information erasure, leading to the identification of the thermodynamic costs associated with finite-time information erasure, known as the finite-time Landauer principle~\cite{berutExperimentalVerificationLandauer2012,zulkowskiOptimalFinitetimeErasure2014,proesmansFiniteTimeLandauerPrinciple2020,zhenUniversalBoundEnergy2021,maMinimalEnergyCost2022}. This field integrates information thermodynamics~\cite{parrondoThermodynamicsInformation2015}—a novel branch of nonequilibrium thermodynamics—with finite-time thermodynamics. In the future, investigating other thermodynamic constraints in information processing~\cite{tasnim2024entropy,zhou2024finite} and optimizing the thermodynamic costs associated with information management are promising directions, which are of great importance for designing energy-efficient high-performance computing and information-assisted thermodynamic cycles~\cite{quan2006maxwell,koskiExperimentalRealizationSzilard2014,zhou2024finite}.

Applying the framework of finite-time thermodynamics to a variety of practical physical processes and systems is also an intriguing area for future research. For instance, the separation of microscopic particles is essential in fields like biology, medicine, and chemical engineering~\cite{kowalczykNanoseparationsStrategiesSize2011,nasiriMicrofluidicBasedApproachesTargeted2020}. The ratchet separation scheme proposed by nonequilibrium thermodynamics offers the advantage of not requiring physical entities, enabling high-performance particle separation~\cite{reimannBrownianMotorsNoisy2002,slapikTunableMassSeparation2019,hermanRatchetBasedIonPumps2023}. Recently, the authors of this paper along with collaborators studied the energy consumption and optimal control of ratchet separation for Brownian particles~\cite{zhaoEngineeringRatchetbasedParticle2024}. 
In addition, the application of finite-time thermodynamic fluctuation relations for efficient free energy estimation remains an active research area~\cite{jarzynskiNonequilibriumEqualityFree1997,crooksEntropyProductionFluctuation1999a,christBasicIngredientsFree2010}, with recent advances in thermodynamic control methods bringing new momentum to the field~\cite{liEquilibriumFreeenergyDifferences2021,whitelamFreeenergyEstimatesNonequilibrium2024,zhongTimeasymmetricFluctuationTheorem2024}. In the nonequilibrium dynamical processes of biological systems, both information processing~\cite{lanEnergySpeedAccuracy2012,sartoriThermodynamicsErrorCorrection2015} and energy utilization~\cite{fangNonequilibriumThermodynamicsCell2020,yangPhysicalBioenergeticsEnergy2021} present opportunities for further research using finite-time thermodynamics. Many of these theoretical issues require experimental validation, and developing diverse platforms to demonstrate nonequilibrium thermodynamic behaviors is also a crucial direction for future experiments in finite-time thermodynamics.

\section{Summary}
We summarize the historical development of finite-time thermodynamics and review the current state of research over the past two decades in this field, focusing on fundamental constraints of finite-time thermodynamic cycles, optimal control and optimization of thermodynamic processes, the operation of unconventional heat engines, and experimental progress. Exploring and utilizing the constraint relations and optimization methods provided by finite-time thermodynamics across different schemes to enhance energy conversion efficiency and reliability is crucial for the new era of global energy transformation and technological revolution. We conclude this paper with three remarks: i) It is necessary to develop new theoretical tools for studying fast-driving thermodynamic processes away from equilibrium; ii) Unconventional heat engines with collective advantages and thermodynamic cycles involving active matters merit further extensive and in-depth research; iii) Integrating finite-time thermodynamics methodologies and paradigms into a broader range of practical thermodynamic tasks and physical systems, such as micro-particle separation, information processing, and battery performance optimization is a promising and challenging development direction in this field. 

\begin{acknowledgments}
Y. H. Ma thanks Wan-Yan Chen and Cong Fu for their careful reading of this manuscript. This work is supported by the National Natural Science Foundation of China for support under grant No. 12305037 and the Fundamental Research Funds for the Central Universities under grant No. 2023NTST017.
\end{acknowledgments}

\bibliography{ref}
\end{document}